\documentclass[preprint,12pt]{elsarticle}

\usepackage{amssymb}
\usepackage{float}
\usepackage{multirow}
\journal{Journal of Crystal Growth}

\begin{document}

\begin{frontmatter}



    \title{MOCVD growth of N-polar GaN on on-axis sapphire substrate: impact of AlN nucleation layer on GaN surface hillock density}


    \author[cnse]{Jonathan Marini}
        \ead{jmarini@albany.edu}
    \author[cnse]{Jeffrey Leathersich}
    \author[cnse]{Isra Mahaboob}
    \author[cnse]{John Bulmer}
    \author[cnse]{Neil Newman}
    \author[cnse]{F. (Shadi) Shahedipour-Sandvik}

    \address[cnse]{Colleges of Nanoscale Science and Engineering, SUNY Polytechnic Institute, Albany NY 12203}

    \begin{abstract}

    We report on the impact of growth conditions on surface hillock density of N-polar GaN grown on nominally on-axis (0001) sapphire substrate by metal organic chemical vapor deposition (MOCVD). Large reduction in hillock density was achieved by implementation of an optimized high temperature AlN nucleation layer and use of indium surfactant in GaN overgrowth. A reduction by more than a factor of five in hillock density from 1000 to 170 hillocks/cm$^{-2}$ was achieved as a result. Crystal quality and surface morphology of the resultant GaN films were characterized by high resolution x-ray diffraction and atomic force microscopy and found to be relatively unaffected by the buffer conditions. It is also shown that the density of smaller surface features is unaffected by AlN buffer conditions.

    \end{abstract}


\end{frontmatter}


\section{Introduction} \label{sec:introduction}

Traditionally, investigation of III-nitride materials and devices have used the Ga-polar orientation $(0001)$. High quality material in this orientation can be obtained using MOCVD on sapphire or SiC substrates. The built-in spontaneous and the stress-induced piezoelectric polarization field in this orientation \cite{Ambacher1999} have been seen to be important properties of the material that can be exploited for advanced device design \cite{Keller2002, Simon2010, Jena2011}. Many devices, including high electron mobility transistors (HEMTs), utilize polarization engineering and take advantage of this field for device behavior and properties that would not be possible in its absence.

The N-polar orientation (000$\overline{1}$) has shown great promise for expanding the existing polarization engineering techniques due to the reversed direction of the spontaneous polarization field. This allows for previously unavailable techniques including improved confinement of channel electrons in N-polar HEMTs as compared to Ga-polar \cite{Rajan2007} and improved extraction of photocarriers in InGaN solar cells \cite{Inoue2010}.

Until recently, development of high quality smooth N-polar GaN films was hampered by the presence of hexagonal hillock surface features \cite{Rouviere1997, Fuke1998, Sumiya1999, Weyher1999}. Hillock formation has been attributed to the nucleation of faster-growing Ga-polar inversion domains in the N-polar matrix \cite{Rouviere1997, Weyher1999}, which has been observed to occur for both the homoepitaxial and heteroepitaxial cases. The lower adatom mobility on the N-polar surface as compared to the Ga-polar surface \cite{Zywietz1998, Jindal2009} means that adatoms are more likely to nucleate at inversion domains and that once formed they cannot be easily overgrown. Additionally, sapphire substrates must be nitridated prior to deposition in order to prepare the surface for N-polar growth \cite{Fuke1998, Sumiya2004} and unoptimized nitridation conditions are shown to lead to higher densities of hillocks \cite{Sun2009}.

The use of misoriented or vicinal substrates has been shown to increase material quality and remove or reduce the presence of hillocks by decreasing the surface terrace width, allowing for adatoms to nucleate at atomic step edges rather than form hillocks \cite{Zauner2000, Zauner2002, Keller2007} and to more easily overgrow those that do form. However, use of vicinal substrates can lead to atomic step bunching and surface roughening \cite{Brown2008}, which is undesirable for devices in which surface properties are an important consideration. Additionally, miscut substrates are more expensive than the traditional nominally on-axis sapphire substrates used for Ga-polar growth, which would increase the cost of N-polar devices as compared to an analagous Ga-polar device.

Some initial work has been done investigating N-polar growth using on-axis substrates. By using optimized nitridation conditions \cite{Sun2009} or indium surfactant \cite{Aisaka2014}, much reduced hillock densities are possible even without the use of vicinal substrates. However, even in these cases the hillocks are not completely suppressed across the surface and there are small-scale surface features present \cite{Polyakov2010}.

Here, we report on the impact of AlN buffer condition and its optimization on further reduction in hillock density in conjunction with indium surfactant. We show that reductions in hillock density can be achieved without compromising crystal quality or surface morphology.

\section{Experiments} \label{sec:experiments}

N-polar GaN growth experiments were carried out in a Veeco D-180 vertical MOCVD system. Growths were done on nominally on-axis c-plane sapphire substrates, with a manufacturer quoted misorientation of 0.2 $\pm$ 0.1$^{\circ}$ towards the m-plane $(1\overline{1}00)$. The substrates were first annealed in H$_2$ environment at 1070$^{\circ}$C for 3 minutes to clean the sapphire surface. Following the anneal, the temperature was ramped up to 1100$^{\circ}$C in preparation for a high temperature (HT) AlN buffer. Prior to the buffer deposition, a 15 second nitridation step was carried under NH$_3$ flow. Next, the high temperature AlN buffer was deposited using Trimethyl-aluminum (TMAl) and NH$_3$ precursors, with TMAl molar flow of 18.5 $\mu$mol/min. The growth conditions of the AlN buffer were varied in order to study the effect on the overgrown N-polar GaN. The parameters investigated were buffer V/III ratio, pressure, and growth time. A summary of the different samples grown using different values in the buffer growth are given in Table~\ref{table:buffer}. Following the HT AlN buffer deposition, a high temperature N-polar GaN film was grown at 1040$^{\circ}$C for 60 minutes using TMGa precursor with molar flow of 80 $\mu$mol/min and V/III ratio of 3000, resulting in a film thickness of about 850nm. During the N-polar GaN growth step, TMIn was also introduced as a surfactant at a flow rate of 40 $\mu$mol/min. Layer thicknesses were measured via \textit{in-situ} reflectance using a k-Space ICE \cite{kspace}. The N-polar GaN surface was studied by optical microscopy and atomic force microscopy (AFM). The resultant crystal quality was characterized via high-resolution x-ray diffraction (HRXRD) using a BedeMatrix-L. The film polarity was confirmed using KOH chemical wet-etching technique which is selective to N-polarity \cite{Guo2013}.

\section{Results and Discussion} \label{sec:results}

In order to study the effect of the buffer growth conditions on both the AlN buffer itself and the overgrown HT GaN morphology, a controlled interruption was performed immediately following the AlN buffer growth.

\subsection{AlN Buffer Layer} \label{subsec:buffer}

\begin{figure}[H]
        \centering
        \includegraphics[scale=0.5]{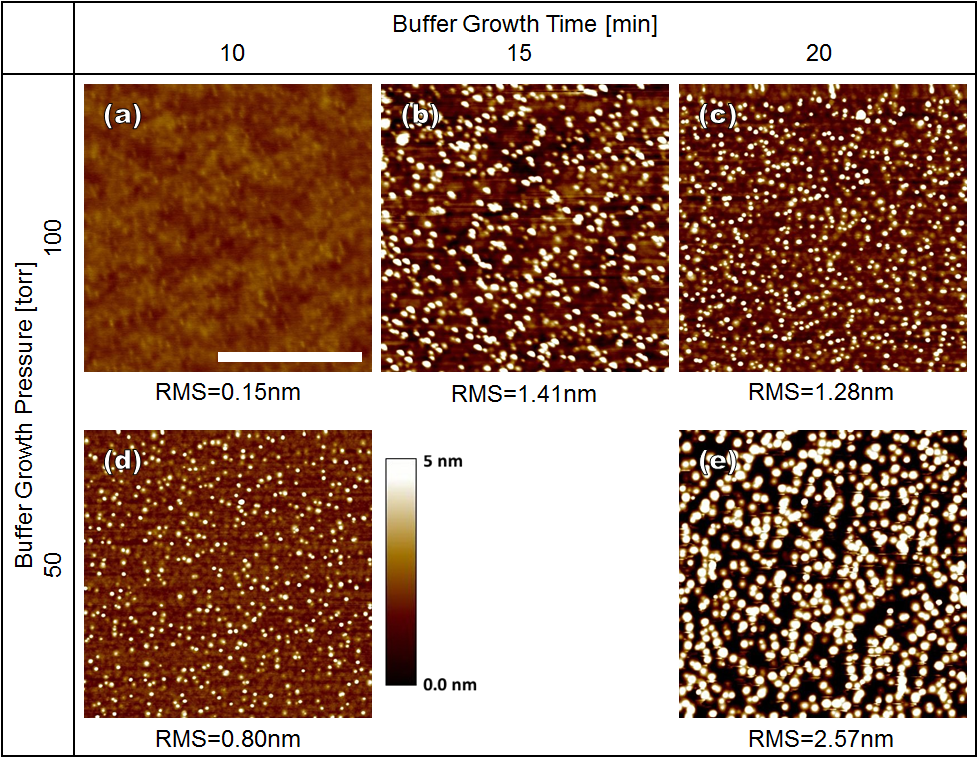}
        \caption{AFM micrographs (2 $\mu$m $\times$ 2 $\mu$m) with inset RMS roughness of the HT AlN buffer at (a) 10 minutes of growth at 100 torr, (b) 15 minutes of growth at 100 torr, (c) 20 minutes of growth at 100 torr, (d) 10 minutes of growth at 50 torr, and (e) 20 minutes of growth at 100 torr. All buffers were grown at low V/III condition (16k) except for (c) which was grown at high V/III condition (24k). Scalebar denotes 1 $\mu$m.}
        \label{fig:afm-buffer}
    \end{figure}

% AFM Buffer Features
Fig.~\ref{fig:afm-buffer}a-e shows AFM micrographs of the as-grown AlN buffer under various growth conditions. The AFM reveals wide variation in the island size and density at the different growth conditions. For the samples grown at a pressure of 50 torr (Fig.~\ref{fig:afm-buffer}d-e), the surface is sparsely covered with small islands after 10 minutes of growth (sample C) and nearing full surface coverage with a high density of islands after 20 minutes of growth (sample A). In contrast, at reactor pressures of 100 torr (Fig.~\ref{fig:afm-buffer}a-c) the growth rate is dramatically lower. After 10 minutes of growth (sample D), there are almost no observable nuclei on the surface, with very minimal island formation being observed, and with 20 minutes of growth (sample B) the island size is similar to 10 minutes of growth at 50 torr, but with increased island density due to the higher growth pressure.

The island density is affected by the growth pressure and V/III ratio. Based on the effect that these parameters have on adatom diffusion length, it can be expected that a higher growth pressure and/or a higher V/III ratio would lead to a higher density of smaller islands \cite{Zhang1997}. Both of these effects are seen in the AFM images of the buffer growths. Fig.~\ref{fig:afm-buffer}b \& \ref{fig:afm-buffer}c shows the effect of V/III ratio, where a higher density of smaller islands is seen in the higher V/III condition (Fig.~\ref{fig:afm-buffer}c). Similarly, Fig.~\ref{fig:afm-buffer}c \& \ref{fig:afm-buffer}d show very similar buffer morphology but with the higher pressure condition showing a higher density of smaller islands.

It has been widely reported that TMAl has parasitic gas-phase reactions with NH$_3$ which results in reduced growth rate and lower Al incorporation in AlGaN films \cite{Chen1996, Mihopoulos1998}. Previous work has shown \cite{Chen1996} that the effect of the parasitic TMAl:NH$_3$ reaction is strongly dependent on growth pressure and temperature. In the present study, at the higher pressure condition (100 torr), the growth pressure combined with the high growth temperature work in tandem to reduce the buffer growth rate by about half as compared to the lower pressure (50 torr) case. Indeed, Fig.~\ref{fig:afm-buffer}a shows that at 10 minutes of growth there is no observable island nucleation in the higher pressure case. To measure island height, a cross sectional cut of Fig.~\ref{fig:afm-buffer}c \& \ref{fig:afm-buffer}d was taken. After 10 min of growth at low pressure and 20 min of growth at high pressure, the AlN islands are ~4-7nm in both cases despite the difference in growth time.

\subsection{N-polar GaN Overgrowth} \label{subsec:ht-gan}

\begin{figure}[H]
        \centering
        \includegraphics[scale=0.5]{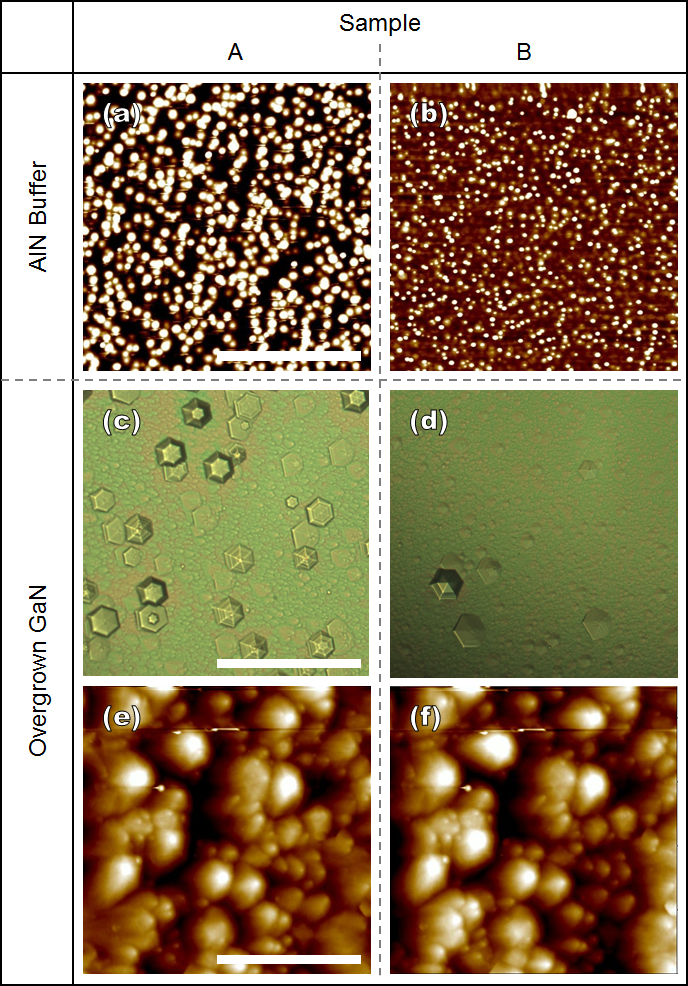}
        \caption{Comparison of samples A and B for both the AlN buffer (a)-(b) and the overgrown GaN (c)-(f). The scalebar on (a) and (b) denotes 1 $\mu$m. Optical microscopy of the surface of the overgrown N-polar GaN is shown in (c) and (d). Hexagonal hillocks can be seen as well as smaller-scale surface features. (c) shows the highest density of hillocks, about 1000 hillocks/cm$^{-2}$ and (d) shows the best sample, with a density of hillocks about 170 hillocks/cm$^{-2}$.  The scalebar on (c) and (d) denotes 1000 $\mu$m. AFM micrographs of the overgrown N-polar GaN away from hillocks showing surface features are shown in (e) and (f), the scalebar denotes 45 $\mu$m.}
        \label{fig:sample-compare}
    \end{figure}

\begin{figure}[H]
        \centering
        \includegraphics[scale=0.4]{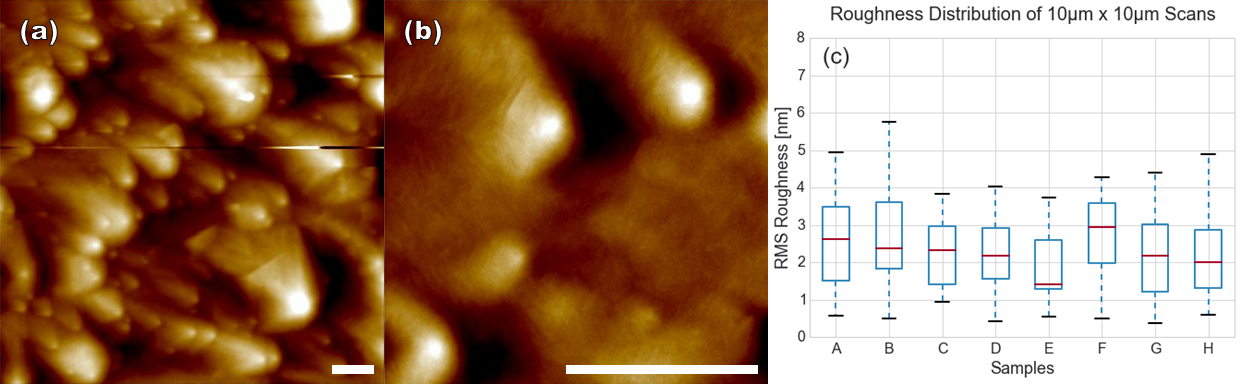}
        \caption{Representative AFM micrographs of surface features on sample E at (a) 90  $\mu$m $\times$ 90 $\mu$m scan size and (b) 20  $\mu$m $\times$ 20 $\mu$m scan size. (c) RMS roughness variation from 10  $\mu$m $\times$ 10 $\mu$m scans for each sample. Scalebar in (a) \& (b) denotes 10 $\mu$m.}
        \label{fig:afm-surface-features}
    \end{figure}

% N-polar Overgrowth
When high temperature N-polar GaN is overgrown on these AlN buffers with the aid of indium surfactant, low hillock densities in the range of 10$^2$-10$^3$ cm$^{-2}$ are observed, which vary depending on the underlying AlN buffer growth conditions, as presented in Fig.~\ref{fig:sample-compare}c-d for samples A and B, respectively. Also visible in the optical images are the smaller surface features seen previously in N-polar growth using on-axis sapphire substrate \cite{Sun2009, Aisaka2014, Polyakov2010}. AFM was performed in regions away from hillocks in order to characterize these features more closely, shown in Fig.~\ref{fig:sample-compare}e-f for samples A and B, and in Fig.~\ref{fig:afm-surface-features}a-b as a representative image from sample E. These features appear at a large density, typically on the order of 10$^5$-10$^6$ cm$^{-2}$ with heights of around 10-20nm. The density and size of these features was seen to be minimally affected by the buffer growth conditions as shown in the box plots of RMS roughness in Fig.~\ref{fig:afm-surface-features}c taken at various points across the wafers. Each box represents the distribution of measured surface roughness over multiple 10$\mu$m $\times$ 10$\mu$m AFM scans across a single sample. The bottom and top of the box represents the first and third quartiles, respectively, of the distribution of the RMS roughness values for measurements at different locations across a particular sample, the line through the box represents the median value, and the ``whiskers'' represent the maximum and minimum values. The similarity in overgrowth surface morphology suggests that the formation condition of these features was not affected by the large differences seen in the island size and density in the AlN buffer. Looking at the samples in Fig.~\ref{fig:sample-compare}, the resultant overgrown GaN surface morphology in Fig.~\ref{fig:sample-compare}e-f shows minimal differences despite large differences in the corresponding buffer morphology and hillock density.

The exact origin of these surface features is unknown at this time, and more investigation will be necessary to determine their formation mechanism and conditions to suppress them. Their formation may possibly be attributed to non-uniformity in the height of AlN buffer nuclei, as some islands were observed to be 2-3 times the height of surrounding islands. The density of these non-uniform spikes roughly corresponds to the density of the observed surface features, suggesting a potential connection. Similar spikes in the nitridated sapphire substrate \cite{Uchida1996} have previously been correlated with hillock formation in N-polar GaN films \cite{Sun2009}. These spikes in the buffer height profile could provide ideal nucleation sites for both hillock formation as well as the smaller surface features. Once these features have formed, even the increased adatom diffusion length due to the indium surfactant may not be enough to overgrow them. Similar surface morphology has also been reported for highly Mg-doped Ga-polar GaN \cite{Ramachandran1999, Fichtenbaum2007} and was attributed to the formation of N-polar inversion domains.

% Hillock Densities
The average hillock density on the samples was quantified by averaging the count of the number of hillocks that appear in several large-scale optical images for each sample. Between the samples with the highest and lowest hillock densities, the density decreases by a factor of 5.9, from around 1000 to 170 hillocks/cm$^{-2}$. Over the rest of the samples studied, the hillock density varies by a factor of 2-3 depending on the buffer growth conditions.

\begin{figure}[H]
        \centering
        \includegraphics[scale=0.5]{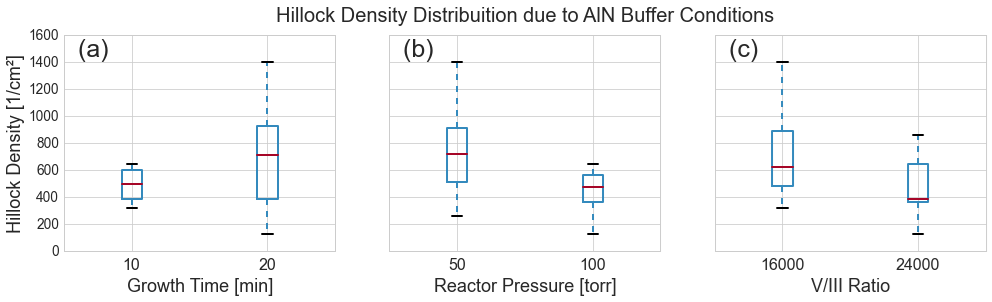}
        \caption{Box plots showing change in hillock density across the sample dependent on (a) growth time, (b) reactor pressure, and (c) V/III ratio for the AlN buffer growth.}
        \label{fig:optical-boxplots}
    \end{figure}

% Effect of Pressure
Fig.~\ref{fig:optical-boxplots} shows the effects of the various AlN buffer growth conditions on hillock density in the overgrown GaN films, presented as a box plot. The growth pressure was observed to have the most significant impact on the hillock density, with the effect being 50\% higher than for other factors. This is consistent with the morphology observed in the AlN buffers, as the growth pressure was the most significant factor among those studied in affecting island formation and growth behavior. The effect of growth time was seen to be highly variable for the 20min case, with both the best and worst samples falling under this condition. This is attributed to the strong interaction effect between growth time and growth pressure. For AlN buffers grown under the higher pressure condition, longer growth times lead to lower hillock densities in overgrown GaN films while at the lower pressure condition the opposite was the case, with lower growth time leading to lower hillock densities. The morphology of the AlN buffers for these two cases is very similar, shown in Fig.~\ref{fig:afm-buffer}a \& \ref{fig:afm-buffer}e, with the island height for both cases being ~4-7nm. In addition to affecting the growth rate, the buffer growth pressure also affects the island density. The effect of island density can be more easily seen by looking at the V/III ratio, which has a similar effect as pressure. Fig.~\ref{fig:optical-boxplots}c shows a strong effect of the buffer V/III ratio, with a higher V/III ratio leading to a lower hillock density. Since the effect of a higher V/III ratio on island size and density is similar to the effect of higher growth pressure, we can attribute some of the improvement in hillock density due to higher pressure to the change in island size and density.

% Island Density & Inversion Domains
It has been observed in both homoepitaxy and heteroepitaxy of N-polar III-nitrides that hillocks originate from Ga-polar inversion domains that are faster-growing than the N-polar crystal surrounding it \cite{Rouviere1997, Weyher1999}. In heteroepitaxy of N-polar AlN on sapphire the formation of these inversion domains has been correlated with voids formed at the sapphire-AlN interface due to decomposition of sapphire in H$_2$ environment \cite{Kumagi2010, Kirste2013, Hussey2014}. An effective method of reducing hillock densities has been to create growth conditions that allow for inversion domains to be overgrown by nucleation at step edges, either via a reduction in terrace width using vicinal substrates or by increasing adatom diffusion length using a surfactant. In the current work, we show that both the effects of buffer pressure and V/III ratio point towards a higher density of islands leading to a reduction in hillock density. A higher density of N-polar AlN islands means adatoms are more likely to nucleate at these islands rather than at voids or other surface features that lead to inversion domain formation. Additionally, a higher density of AlN islands also reduces the time to film coalescence which allows for the voids to be grown over more quickly and reduce the chance of inversion domain propagation through the film.

% Crystal Quality
The resultant crystal quality of the N-polar GaN films was determined by HRXRD measurement, which can be used to evaluate dislocation density \cite{Ayers1994}. The full-width at half-maximum (FWHM) of the x-ray rocking curve was evaluated at the (0002) and (10$\overline{1}$5) planes. Table~\ref{table:xrd} shows the measured FWHM of the overgrown N-polar GaN corresponding to the AlN buffers shown in Fig.~\ref{fig:afm-buffer} (samples A-D). Across all evaluated samples (A-H) the variation in measured FWHM was small for both planes. The coefficient of variation ($c_v = \sigma / \mu$) was calculated as a magnitude-independent evaluation of the variability in FWHM, and was 11\% ($\mu = 389, \sigma = 43.98$) for the (0002) plane and 4.8\% ($\mu = 521, \sigma = 24.95$) for the (10$\overline{1}$5) plane. Even though the hillock density was seen to change by as much as 5-6 times across the various buffer conditions, the crystal quality of the resultant GaN was only seen to vary by 11\% or less. Although we do not observe correlation between hillock density and dislocation density, previous studies have shown that clusters of defects with a higher density than the surrounding material can be seen at the apex of hillocks \cite{Zhou2013}.

\section{Conclusions} \label{sec:conclusions}

It is found that optimized AlN buffer growth conditions can reduce the hillock density in N-polar GaN films by more than a factor of five in conjunction with the use of indium surfactant. This reduction in hillock density was found to be independent of changes to crystal quality and local surface morphology, with both qualities undergoing minimal change with major differences in buffer morphology and surface coverage. It was found that higher island density in the AlN buffer results in reduced hillock density in overgrown films. Small-scale surface features with heights of 10-20nm were found on N-polar GaN films which is characteristic of reports for N-polar growth using on-axis substrates. These surface features may be a result of height non-uniformities in the AlN nuclei which manifests as islands that are taller than the surrounding growth. This correlation is consistent with previously reported results of protrusions resulting from unoptimized nitridation conditions.

\section*{Acknowledgements}

This work was supported by a grant (\#1473194) from the National Aeronautics and Space Administration.


\newpage

\bibliographystyle{elsarticle-num}
\bibliography{npolar_doe}

\newpage

    \begin{table}[H]
        \centering
        \begin{tabular}{c|ccc}
            \multirow{2}{*}{Sample} & \multicolumn{3}{c}{AlN Buffer Growth Parameters} \\
            \cline{2-4} & Time [min] & Pressure [torr] & V/III Ratio \\
            \hline
            A & 20 &  50 & 16000 \\
            B & 20 & 100 & 24000 \\
            C & 10 &  50 & 16000 \\
            D & 10 & 100 & 16000 \\
            E & 20 & 100 & 16000 \\
            F & 10 & 100 & 24000 \\
            G & 20 &  50 & 24000 \\
            H & 10 &  50 & 24000
        \end{tabular}
        \caption{AlN buffer growth parameters for each sample investigated in the current study.}
        \label{table:buffer}
    \end{table}

    \begin{table}[H]
        \centering
        \begin{tabular}{c|cc}
            \multirow{2}{*}{Sample} & \multicolumn{2}{c}{FWHM of XRCs [sec$^{-1}$]} \\
            & (0002) & (10$\overline{1}$5) \\
            \hline
        % Sample & \multicolumn{2}{c|}{Buffer Growth Parameters} & \multicolumn{2}{c}{FWHM of XRCs [sec$^{-1}$]} \\
        % & Pressure & Growth Time & (0002) & (10$\overline{1}$5) \\
        % \hline
        A & 395 & 532 \\
        B & 396 & 515 \\
        C & 344 & 513 \\
        D & 429 & 550
        \end{tabular}
        \caption{Crystal quality of overgrown N-polar GaN for samples A-D as measured by HRXRD.}
        \label{table:xrd}
    \end{table}

\end{document}